*Article*

# AI Across Borders: Exploring Perceptions and Interactions in Higher Education

**Juliana Gerard [1,*], Sahajpreet Singh [2], Morgan Macleod [1], Michael McKay [3], Antoine Rivoire [4], Tanmoy Chakraborty [2,5] and Muskaan Singh [6]**

1 School of Communication and Media, Ulster University, Belfast BT1 6AG, Northern Ireland, UK; m.macleod@ulster.ac.uk
2 Department of Electrical Engineering, Indian Institute of Technology Delhi, New Delhi 110016, India; sahaj.phy@gmail.com (S.S.); tanchak@iitd.ac.in (T.C.)
3 School of Medicine, Ulster University, Derry (Londonderry) BT15 1ED, Northern Ireland, UK; m.mckay1@ulster.ac.uk
4 Centre for Digital Learning Enhancement, Ulster University, Belfast BT1 6AG, Northern Ireland, UK; a.rivoire@ulster.ac.uk
5 Yardi School of Artificial Intelligence, Indian Institute of Technology Delhi, New Delhi 110016, India
6 School of Computing, Engineering and Intelligent Systems, Ulster University, Derry (Londonderry) BT15 1ED, Northern Ireland, UK; m.singh@ulster.ac.uk
* Correspondence: j.gerard@ulster.ac.uk

**Abstract**

This study investigates students' perceptions of Generative Artificial Intelligence (GenAI), with a focus on Higher Education institutions in Northern Ireland and India. We collect quantitative Likert ratings and qualitative comments from 1211 students on their awareness and perceptions of AI and investigate variations in attitudes toward AI across institutions and subject areas, as well as interactions between these variables with demographic variables (focusing on gender). We found the following: (a) while perceptions varied across institutions, responses for Computer Sciences students were similar, both in terms of topics and degree of positivity; and (b) after controlling for institution and subject area, we observed no effect of gender. These results are consistent with previous studies, which find that students' perceptions are predicted by prior experience; crucially, however, the results of this study contribute to the literature by identifying important interactions between key factors that can influence experience, revealing a more nuanced picture of students' perceptions and the role of experience. We consider the implications of these relations, and further considerations for the role of experience.

## 1. Introduction

With the increasing presence of Generative AI (GenAI) throughout Higher Education (HE), many organizations involved in HE have produced statements in response to the emergent issues posed by the use of GenAI. In addition to the numerous documents released by individual institutions regarding their policies, joint policy statements and recommendations have been issued by bodies such as the Russell Group in the UK (Russell Group, 2023), the Group of Eight universities in Australia (Group of Eight, 2023), and the European Network for Academic Integrity (ENAI) (Foltynek et al., 2023). Additionally, since 2021, an annual report on AI and its role in the UK's tertiary education system has been issued by the Joint Information Systems Committee (JISC) (JISC, 2023, 2024, 2025).

In general, these documents reflect several common themes. While expressing concern regarding possible inaccuracies and biases in AI-generated content, as well as the potential misuse of AI tools for acts such as plagiarism, the potential benefits of GenAI are also acknowledged. Emphasis is placed on the importance of providing support and training for staff and students in the use of AI, fostering collaboration for the exchange of best practices, and overcoming inequalities and barriers that may impede some students' access to GenAI.

These themes have been identified based on stakeholder consultations across HE, which take various forms. For example, concerns about GenAI in HE contexts may be voiced within focus groups, written statements, or qualitative interviews. In addition, more directed questionnaires may be designed to measure specific aspects of these concerns on a larger scale with quantitative measures (e.g., Likert ratings). These qualitative and quantitative approaches have been taken across a range of studies that have aimed to understand concerns related to GenAI in HE. These studies are largely descriptive in nature, focusing generally on one population as a whole (e.g., UK students), based on a common measure, either qualitative or quantitative. In addition, previous comparative approaches have investigated the use of GenAI based on factors that varied within the population, or across institutions.

In this study, we expand on these approaches to explore student perceptions of GenAI across subject fields/courses of study, for both qualitative and quantitative measures: we compare these perceptions (a) across international contexts, in institutions in Northern Ireland and India, and (b) across subject areas. Importantly, these two factors can co-vary with the demographic factors that have previously been observed to predict perceptions, and can impact on prior experience. We find that both factors predict variation in perceptions—across institutions and subject fields, in both qualitative and quantitative measures. This variation has implications for the interpretation of variation in previous studies, and raises further practical considerations for GenAI in HE—particularly in relation to prior experience.

The remaining sections are laid out as follows: in the next section, we review previous studies on perceptions of GenAI in HE in more detail, focusing on previous methodological approaches and predicting factors; following this, we present our study methods and design.[1] We then present our results, including an analysis of our qualitative data, followed by a comparison of the qualitative and quantitative data, and finally our quantitative data analysis. In the final section, we reflect on these results and conclude the paper.

## 2. Literature Review

### *2.1. AI Perceptions: Methodological Approaches*

#### 2.1.1. Quantitative Approaches

Previous quantitative studies on student perceptions of AI in education have observed variations in perceptions across a range of factors. These include individual student factors (e.g., demographics) and subject area, as well as variation based on other AI-related factors, particularly prior experience with GenAI. Many of these quantitative findings are also complemented by qualitative data, providing further insight into the sources of this variation (Amani et al., 2023; Smolansky et al., 2023).

#### 2.1.2. Qualitative Analyses

Previous qualitative research on perceptions of AI has adopted a variety of methodologies. The UK Government made use of a multi-stage process to solicit responses from different stakeholders about regulatory approaches to AI, beginning with open questions on broad topics (Department for Digital, Culture, Media & Sport, 2022) and then moving to a greater number of open and closed questions on more specific issues (Department for Science, Innovation & Technology, 2023). Following analysis of the responses, the results were used to inform recommendations for future policy and legislation. JISC has conducted a series of in-person focus groups with students from multiple institutions. The comments from these focus groups were then analyzed, and from the results, common themes were identified regarding student concerns and needs, as well as the ways in which students make use of AI (JISC, 2024). At Edinburgh Napier University, the 'ChatGPT & Me' dataset was produced; this used an online Padlet to collect anonymous free-form posts, in order to provide data for research on students' attitudes toward GenAI (Drumm et al., 2023). This approach has since been repeated to produce a longitudinal dataset from the same population, as well as a further dataset from Ulster University.

All of these studies have focused exclusively on people living or studying in the UK; many studies from other countries are similarly national in focus (e.g., Laï et al., 2020; Lin et al., 2023; Neher et al., 2023). Less work has been carried out to compare qualitative responses across a broader, transnational set of data, although some studies that have included both qualitative and quantitative components have taken more comparative approaches.

### *2.2. AI Perceptions: Predicting Factors*

Perceptions of AI in education have been observed to vary depending on different factors that vary within student populations. For example, both perceptions and use of AI vary by students' first language, with students who are non-native English speakers exhibiting more positive views than native speakers (Baek et al., 2024; Warschauer et al., 2023; Shaikh et al., 2023), although with some variation depending on the study location (Kelly et al., 2023); such variation has been interpreted in terms of the potential of AI to support tasks related specifically to language learning (e.g., Baek et al., 2024).

Predicting factors have also included students' subject areas (Baek et al., 2024; Petricini et al., 2024; Kelly et al., 2023), as well as demographic factors like age and gender (Baek et al., 2024; Kelly et al., 2023) and prior experience with GenAI (Faruk et al., 2023; Amoozadeh et al., 2024; Baek et al., 2024). Crucially, these factors can tend to co-vary; thus, while perceptions may be predicted by demographic factors, these factors alone are unlikely to *cause* the observed variation. This consideration is key for the development of institutional recommendations, based on these factors. Variation by experience is also reflected in the analyses of further qualitative data, both in terms of sentiment and by topic (Petricini et al., 2024; see also Kelly et al., 2023).

This variation by prior experience adds crucial context to the other factors: as a new technology that is rapidly changing the landscape of higher education, GenAI has influenced the student experience across these factors to varying degrees. However, this influence has changed since the initial release of ChatGPT in November 2022 by OpenAI and will continue to change with the increasing adoption of AI across the sector (Li et al., 2024; JISC, 2024). Therefore, while the other factors are largely constant, this is not likely the case for prior experience—more so with most quantitative studies to date reporting student perceptions from 2023. Therefore, with changes in the influence of GenAI throughout education—at all levels—we expect changes in students' experience with AI, which in turn may change their perceptions.

In addition to timing, further considerations are also required for the study location: most studies focus on one student population, primarily in US contexts. A comparative approach across international contexts will allow for broader generalizations across populations. We take this approach in the current study, which builds on previous studies to achieve a more nuanced understanding of students' perceptions of AI, and implications for adoption in HE.

### 2.3. The Current Study

In this study, we investigate students' perceptions of GenAI in HE, as well as their general awareness of AI. We aim to explore variation in perceptions, and in particular we ask the following question: which factors predict variation in perceptions, and do these factors interact with each other? As reviewed in Section 2.2, previous studies have observed that variation is predicted by demographic factors; however, we hypothesize that (a) this variation can also be explained by other factors—particularly those which reflect variation in experience (Petricini et al., 2024)—and (b) these other factors may additionally interact with each other, reflecting a more nuanced picture for why students' perceptions may vary in the first place. In particular, we investigate variation across HE institutions and across subject areas, and how these factors interact with demographic variables. We focus on institutions outside the more commonly studied US context, looking instead at Northern Ireland (Ulster University) and India (the Indian Institute of Technology Delhi). Notably, while guidelines for GenAI use have been released at a national level in the UK and at numerous UK institutions, neither were the case for India at the time of data collection; IIT Delhi has recently released guidelines and is the first HE institution in India to do so (Agarwal et al., 2024; PTI, 2025). We use a combination of quantitative Likert ratings and qualitative free-response comments, and compare qualitative and quantitative responses first across institutions, and next across subject areas. Finally, we explore variation based on demographic variables, which also predicted students' perceptions in previous studies.

Crucially, we find variation across all factors—institution, subject area, and demographics; however, a more nuanced picture emerges when we consider interactions between these factors, which can account for uneven distributions across conditions. This has further implications for the role of experience with GenAI, and the interpretation of these effects in previous studies.

## 3. Methods

Students' responses were collected with a questionnaire, which was adapted from the protocol developed by Petricini et al. (2024). The questionnaire included quantitative measures of students' awareness and perceptions of AI, followed by the option to enter a text comment which served as a qualitative measure. In the following sections, we describe each measure in turn, and the motivations for these adaptations.

### 3.1. Participants

The participants were undergraduate or postgraduate students at Ulster University (UU; N = 512) or undergraduate or postgraduate students at the Indian Institute of Technology Delhi (IITD; N = 699), for a total of 1211 participants across both institutions. Most participants were undergraduates (N = 948), and the results presented in Section 4 do not change when postgraduate students were removed from the analysis. We therefore include both within the single category of "students" for the purposes of this study.[2]

All participants responded to the quantitative sections of the questionnaire. Quantitative responses were provided for the full sample, while a subset also entered a qualitative comment (UU N = 192, IITD N = 235). Most participants were aged 18–24 years or 25–39 years, with some variation in the participant distribution across institutions for both age (Table 1) and gender (Table 2). We revisit this variation in the context of academic field below.

**Table 1.** Participant age, by institution.

| University | Age | | | | | Total |
|---|---|---|---|---|---|---|
| | 18–24 | 25–39 | 40–59 | 60+ | N/A | |
| Ulster University | 272 | 178 | 56 | 5 | 1 | 512 |
| IIT Delhi | 671 | 24 | 0 | 1 | 3 | 699 |
| Total | 943 | 202 | 56 | 6 | 4 | 1211 |

**Table 2.** Participant gender, by institution.

| University | Age | | | | | Total |
|---|---|---|---|---|---|---|
| | Male | Female | Non-Binary | Other | N/A | |
| Ulster University | 201 | 292 | 10 | 4 | 5 | 512 |
| IIT Delhi | 563 | 132 | 2 | 0 | 2 | 699 |
| Total | 764 | 424 | 12 | 4 | 7 | 1211 |

To be eligible for the study, participants needed to be enrolled in an undergraduate or postgraduate course at UU or IITD. They were recruited via university listserv in the spring (UU) and summer (IITD) of 2024, and all participants received a GBP 10 (UU) or INR 500 (IITD) voucher for their participation. The study received ethics approval from Ulster University, and participants indicated consent to participate before proceeding to the study.

### 3.2. Design and Procedure

The questionnaire for both institutions included four main components. The first two of these involved rating a series of statements about awareness and perceptions of AI, respectively. Participants then provided demographic information and finally had the option to enter a qualitative comment. In the following sections, we provide further detail on each of these components, before reviewing the planned analysis.

#### 3.2.1. Awareness and Perceptions

In the first two sections of the questionnaire, participants rated statements relating to their awareness of AI, followed by statements about their perceptions of AI. The statements were rated on a 5-point scale from "Strongly disagree" to "Strongly agree", and the full set of statements in these sections is presented in Table 3.

In the first section, students considered statements which related both to general awareness of AI (A1–A3 in Table 3) and to AI in education contexts (A4–A8 in Table 3). The awareness questions were intended to gauge students' general familiarity with AI—independently of their course content—as well as any course-specific content which may have been available at the time.

Next, participants rated statements related to their perceptions of AI in academia. These statements ranged from perceptions of the students' use of AI (P1–P2, P4 in Table 3), instructors' use of AI (P6–P9 in Table 3) and the broader context for AI in education (P3, P5 and P10 in Table 3).

For the quantitative analysis, ratings on the 5-point scale from "Strongly disagree" to "Strongly agree" were mapped to corresponding numeric values from 1 to 5, respectively (apart from the three exceptions noted in Table 3).

**Table 3.** Statements about AI awareness (A) and perceptions (P), rated from "Strongly agree" to "Strongly disagree".

| Statement |
|---|
| A1. I am familiar with the concept of artificial intelligence (AI) |
| A2. I am familiar with ChatGPT or other AI text generation tools |
| A3. I have experience using ChatGPT or other text generation tools |
| A4. My instructors have addressed the use of AI (especially ChatGPT and other text and image generation tools) in my modules |
| A5. My instructors have integrated AI generators like ChatGPT into their instruction |
| A6. I plan to use ChatGPT or similar tools for my coursework in the future |
| A7. I have received instructions about how to use ChatGPT or similar tools |
| A8. I would be open to receiving instructions about how to use ChatGPT or similar tools |
| P1. Students' use of AI text generation tools to complete coursework is prevalent in higher education |
| P2. Students' use of AI text generation tools to complete coursework is inevitable |
| P3. Artificial Intelligence has value in education |
| P4. Students should be restricted from using AI for coursework [a] |
| P5. AI is used in education for good and helpful reasons |
| P6. Instructors are confident in their use of AI in academic settings |
| P7. I would feel confident knowing an instructor was using an AI-created teaching resource |
| P8. I would want to be informed if my instructor was using AI-created resources on courses [a] |
| P9. I trust AI in marking my assignments and assessments for my modules instead of my instructor |
| P10. Use of AI text generation tools to complete coursework is inconsistent with academic integrity policies at the University [a] |

[a] The scoring of this question was reversed for the quantitative analyses.

#### 3.2.2. Demographic Information

After indicating their agreement with the statements in the first two sections, participants were asked to complete a series of demographic questions, including their current course of study. From their selected course, we categorized participants into broader subject areas. This classification was based on Clarivate's Web of Science classification system (Clarivate Analytics, 2020), which groups subjects into the high-level categories of Arts & Humanities, Life Sciences & Biomedicine, Physical Sciences, Social Sciences, and Technology. For analysis, we separated Computer Sciences from the remaining Technology categories, as Computer Sciences students may be more closely involved with AI in the course of their studies than students from other Technology fields such as metallurgy. For students on multi-subject courses (e.g., English with History), only the major subject was counted. Some students declined to provide information on their specific course, but chose to give information on their faculty and/or department; where possible, they were assigned a subject category based on this information. A separate field was used to track students whose subject area was imputed in this way; however, these students have not been excluded from the analysis presented here, as the categories used are broad enough for the imputed information to be reliable.

The demographics questions were selected based on the objectives outlined above. In particular, we aimed to tease apart demographics that tend to vary across subject areas and institutions from variation in responses by subject area or institution. That is, for any course that may over-represent a particular demographic profile (e.g., by gender, age, etc.), variation by course would also appear as variations by this same profile.

For example, our sample has a skewed distribution for gender by both course and institution (Table 4); this variable is represented in different proportions across courses at UU (e.g., with more females in Social Sciences, but more males in Tech), and also in our overall sample across institutions (more females in the UU sample and more males in the IITD sample). We therefore avoid this confound with gender by querying the participants' courses in addition to other demographic information.

**Table 4.** Student distribution by gender and subject area across universities, not including Non-binary and Other due to insufficient data.

| University | Gender | Arts | Bio Sci | Phys Sci | Soc Sci | Comp Sci | Tech | Total |
|---|---|---|---|---|---|---|---|---|
| UU | Female | 32 | 82 | -- | 145 | 13 | 18 | 290 |
| | Male | 19 | 25 | -- | 97 | 31 | 27 | 199 |
| IITD | Female | -- | 9 | 6 | 9 | 31 | 77 | 132 |
| | Male | -- | 27 | 31 | 3 | 160 | 341 | 562 |

Demographic questions were based on census categories in Northern Ireland for UU and in India for IITD. The full set of demographic questions is presented in Appendix A. For the quantitative analysis in the Results section, we focus on the gender variable as a case study in covariation by course and institution and return to the remaining demographics and implications in the discussion section.

#### 3.2.3. Qualitative Comments

After completing the demographics section, participants had the option to enter a free text comment, with the following prompt:

> *(Optional) "Please enter any further comments about this questionnaire, your experience with AI in academia, or anything else that you would like us to consider".*

No further instructions were provided, and participants were free to expand on their ratings, comment further on other aspects of AI, or provide any other relevant information.

As comments were optional, not all participants included a free text response for this field; however, many did include one: there were comments from 427 participants in total, with 192 comments from UU and 235 comments from IITD. This was sufficient for a full qualitative analysis across institutions and subject areas (see Table 6), presented in the following sections.

## 4. Results

To analyze the students' perceptions of AI in education contexts, we consider both the quantitative and qualitative response types. We start by considering the qualitative comments, first in isolation, and next alongside the quantitative ratings. As the comments were not limited beyond the general theme of AI, they provided various forms of information. We focus first on the sentiment of each comment—positive, negative, or neutral. However, considering a comment on perception may include both positive and negative views combined about different aspects of AI, a single sentiment tag for the entire comment is insufficient. We explore the aspect-based sentiment analysis (ABSA) approach to gain a better understanding of the comments. ABSA is helpful in two ways: (1) extraction of topical sentiments and (2) usefulness of extracted aspects to look at the topical coverage.

With a numerical sentiment score for each comment, we next compare the qualitative comments and quantitative responses directly, via a correlation analysis. This direct comparison allows for a quality check: given that the same participant produced both scores, these two measures (i.e., qualitative comments and quantitative ratings) should extract similar perceptions. We test this prediction with the correlation analysis.

Finally, we focus on the full set of quantitative responses (including from those who did not enter a comment). We conduct an Exploratory Factor Analysis with questions on awareness and perceptions of AI, followed by regression analyses by institution, subject area, and the demographic variable gender.

### *4.1. Qualitative Data: Sentiment Analysis*

On average, comments across all responses consisted of 45 words (SD = 38 words). These averages differed somewhat across institutions (with an average of 56 (SD = 42) words for UU and 36 (SD = 32) for IITD), and across subject areas (Table 5). These initial differences in word counts hint at further variation across both variables, i.e., institution and subject area. Thus, we next explore the content of these comments, starting with a sentiment analysis.

**Table 5.** Descriptive statistics of participants' comments.

| Word Count | Arts | Bio Sci | Phys Sci | Soc Sci | Tech | Comp Sci | Overall |
|---|---|---|---|---|---|---|---|
| **Average (SD)** | 81 (57) | 48 (36) | 29 (26) | 50 (37) | 41 (37) | 38 (31) | 45 (38) |

For sentiment tagging, we utilized a RoBERTa-based sentiment analysis model (Loureiro et al., 2022)[3], and observed generally more positive sentiment scores across comments from IITD compared to UU (Table 6). This suggests that IITD participants view AI as a beneficial tool in their regular educational activities, whereas those at UU are more neutral.

**Table 6.** Overall sentiment classification statistics of participants' comments concerning different categories, i.e., institutions and fields of study, with totals in bold.

| Category | Sub-Category | # Comments | Sentiment Count | | |
|---|---|---|---|---|---|
| | | | # Positive | # Negative | # Neutral |
| **Institutions** | Ulster University | 192 | 50 | 56 | 86 |
| | IIT Delhi | 235 | 99 | 41 | 95 |
| | **Total** | **427** | **149** | **97** | **181** |
| **Fields** | Arts | 22 | 7 | 12 | 3 |
| | Biological Sciences | 58 | 18 | 10 | 30 |
| | Physical Sciences | 13 | 7 | 3 | 3 |
| | Social Sciences | 100 | 25 | 30 | 45 |
| | Technological Sciences | 152 | 65 | 33 | 54 |
| | Computer Sciences | 81 | 27 | 9 | 45 |
| | **Total** | **426** | **149** | **97** | **180** |

While this contrast captures an overall difference in sentiment, it fails to explain the aspects that are responsible for the respective overall sentiments. To address this limitation, we used the SetFit-based (Tunstall et al., 2022) aspect-based sentiment analysis (ABSA) model[4], which identifies the topics/aspects mentioned in each comment and their associated sentiments. This approach enables a better understanding of the range of perspectives expressed by the participants. Figure 1 shows the results from ABSA extension, revealing key topics/aspects associated across sentiments and institutions.[5]

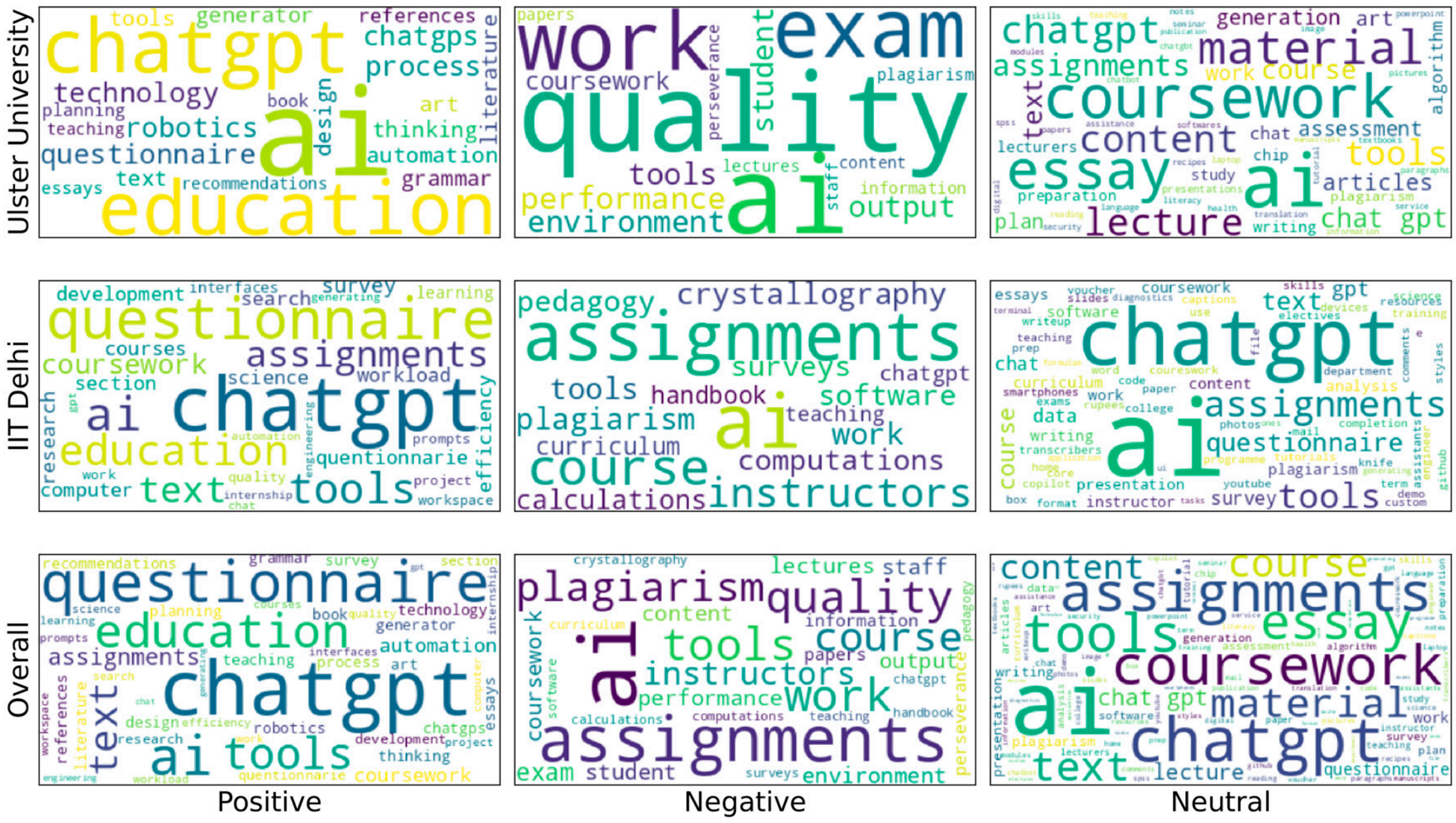

**Figure 1.** Word cloud for aspect-based sentiment analysis (institutions).

For instance, "chatgpt" and "education" featured prominently in positive aspects across both institutions, while keywords observed for positive aspects across both institutions. Meanwhile, less overlap was observed for frequent terms in negative sentiments, with negative comments at UU featuring words like "quality", "work" and "exam", while IITD negative comments were more varied, often mentioning "assignments". To better understand these variations in perception, we next explored these keywords through the lens of academic fields (Arts, Biological Sciences, Physical Sciences, Social Sciences, Technological Sciences, and Computer Sciences), in Figure 2.

Note from Table 6, comments vary across the different subject areas in their distribution across positive, negative and neutral sentiments. Meanwhile, a large ratio of comments from IITD come from participants with technical education background, while UU has a higher number of social science participants (Table 4). Thus, keywords like "questionnaire" and "assignments" (refer to Figures 1 and 2) which appear frequently at IITD may be due to a contrast between institutions, given IITD's technical focus, as they align closely with technical coursework; however, these keywords may also reflect the skewed distribution across academic fields. Similarly, the alignment of UU's perceptions toward "education", "quality", "performance", "essay", and "environment" (see Figures 1 and 2) may reflect an institution-wide focus from the arts and social science community, where writing and diverse ideology play the most vital role; these keywords may also reflect the larger proportion of these fields within the UU sample.

If so, then we expect to observe corresponding patterns between the qualitative and quantitative data, and also the same contrasts in the quantitative data. In the following sections, we first assess the correspondence between the qualitative and quantitative elements, followed by an analysis of the quantitative data to confirm the second prediction.

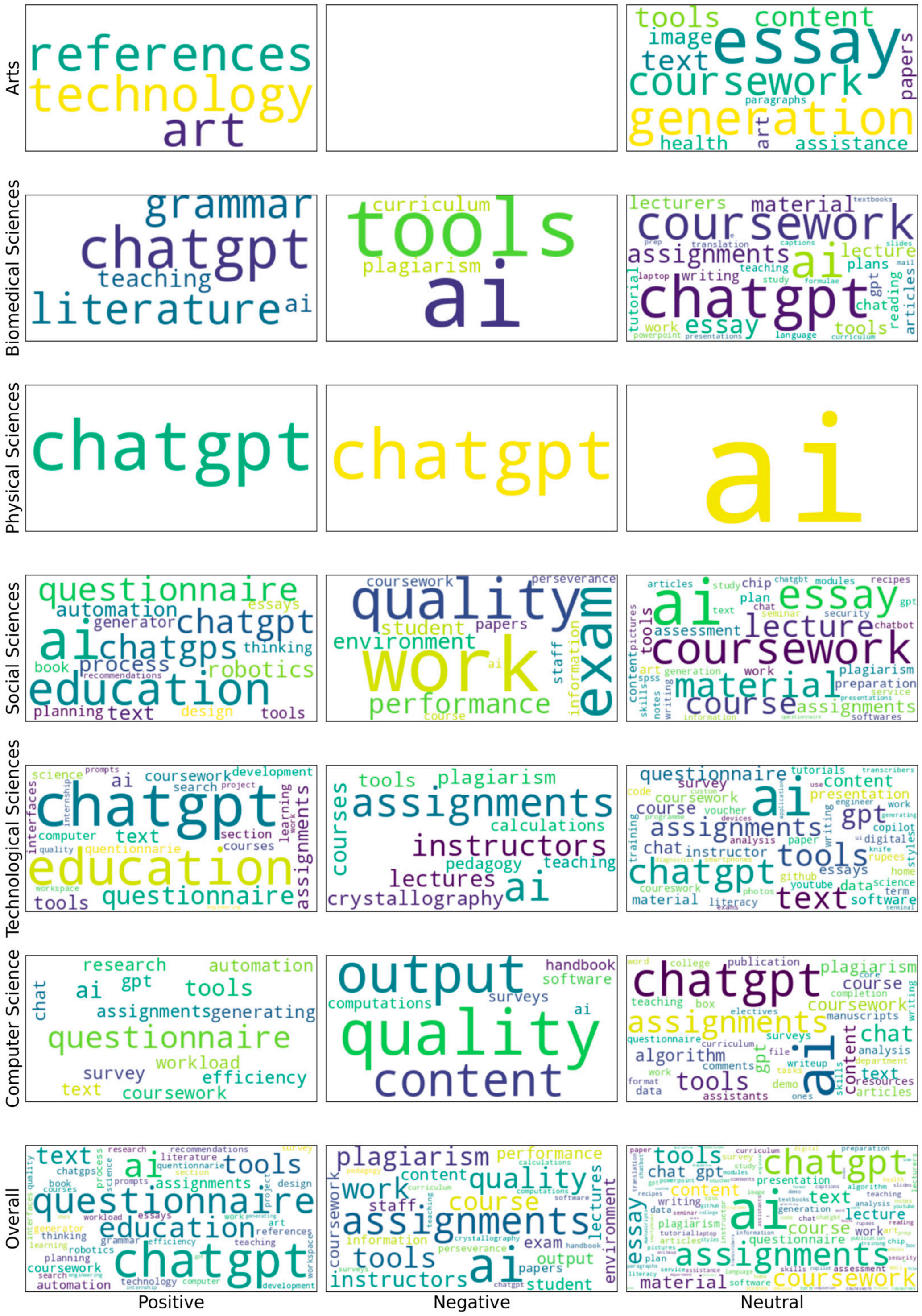

**Figure 2.** Word cloud for aspect-based sentiment analysis (fields).

## 4.2. Comparison Between Qualitative and Quantitative Data

Meanwhile, to explore the relations between qualitative responses and textual comments, we encoded positive, neutral, and negative sentiments to 5, 2.5, and 0, respectively. Finally, we analyzed the relationships between question pairs and textual comments by conducting correlation experiments to assess the extent to which perception scores are interrelated, in Figure 3.

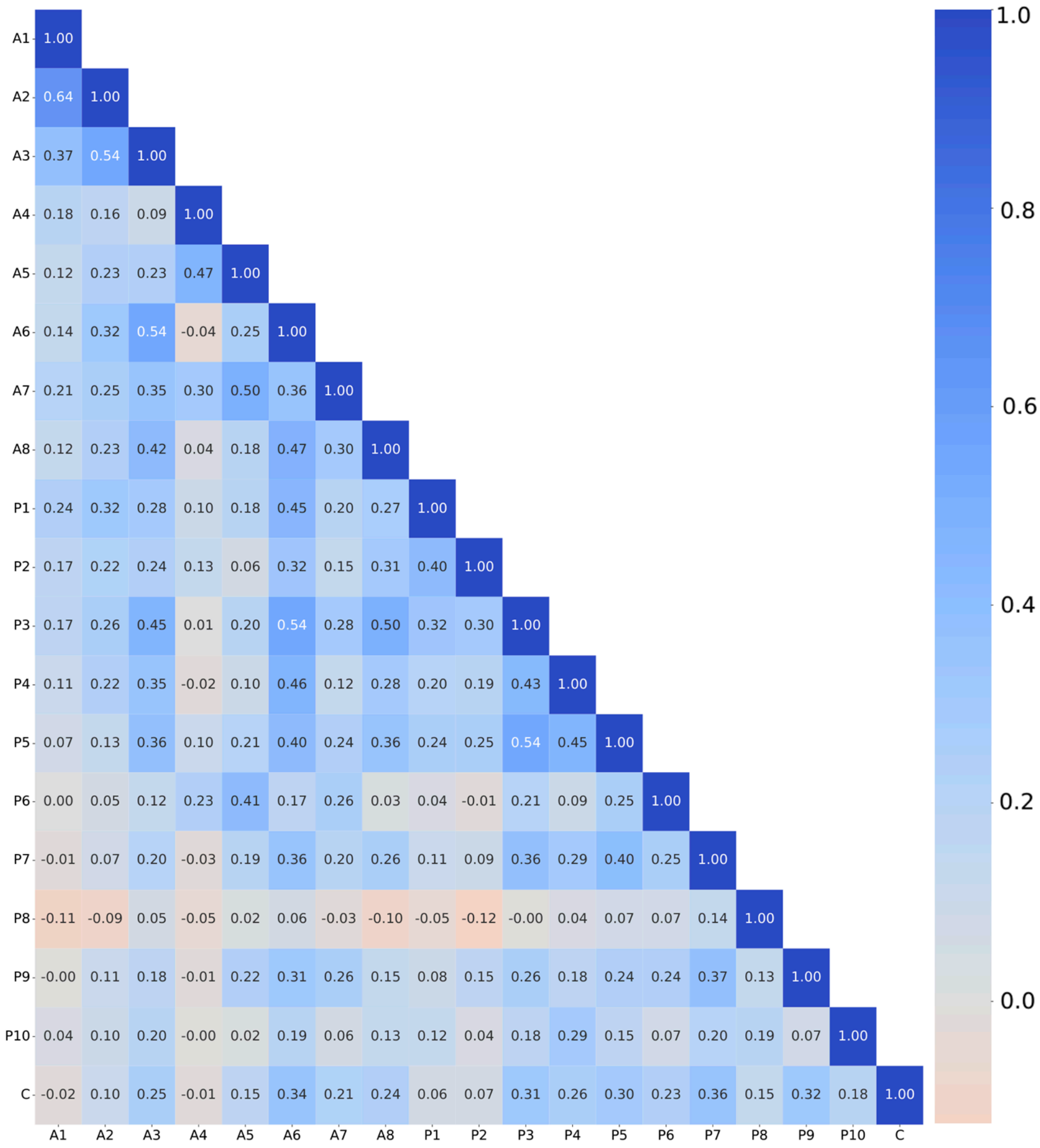

**Figure 3.** Correlation heatmap for different question pairs (A1–P10) and qualitative comments' sentiments (C).

We focus first on the correlations between the individual questions and comment sentiments, which are presented in the final row of the heatmap in Figure 3. This final row reflects the correlations between the participants' responses to survey questions and sentiments for their qualitative comments—importantly, these relations vary considerably. For example, the strongest relations with sentiment scores were observed for ratings to A6 (*I plan to use ChatGPT or similar tools for my coursework in the future*) and P7 (*I would feel confident knowing an instructor was using an AI-created teaching resource*). These questions address different aspects of AI use in education, with the former related to personal experience and the latter regarding perceptions. Correlations for the remaining questions varied widely, with weak or no relation both for questions on awareness and perceptions. This suggests that the content of the comments themselves was highly variable, which is consistent with the wide range of topics identified by the ABSA in the previous section.

Meanwhile, stronger correlations are observed in Figure 3 between many of the questions themselves—both within each section (on awareness and perceptions, respectively) and across these sections. This suggests that many of these questions tap into common factors that may contribute to participants' responses. We investigate these potential factors next via a factor analysis.

### 4.3. Quantitative Data: Comparison Across Fields and Institutions

As described above, participants' ratings of statements about awareness and perceptions of AI were adapted to a 1–5 scale, with 1 corresponding to "Strongly disagree" and 5 to "Strongly agree". The mean ratings based on this adaptation are presented for each question in Tables 7 and 8. Of note from these preliminary results is that in all but one case (A4), the mean score for IITD was higher than that for UU. As expected, the highest scores are observed for A1, for general familiarity with AI. However, to investigate these contrasts further, we conducted an Exploratory Factor Analysis (EFA).

**Table 7.** Mean and standard deviation of AI awareness-related statements for UU, IITD, and overall.

| Statement | UU | IITD | Overall |
|---|---|---|---|
| A1. I am familiar with the concept of artificial intelligence (AI) | 4.39 (0.74) | 4.49 (0.67) | 4.45 (0.70) |
| A2. I am familiar with ChatGPT or other AI text generation tools | 4.32 (0.86) | 4.66 (0.56) | 4.52 (0.72) |
| A3. I have experience using ChatGPT or other text generation tools | 3.71 (1.31) | 4.71 (0.57) | 4.29 (1.07) |
| A4. My instructors have addressed the use of AI (especially ChatGPT and other text and image generation tools) in my modules | 3.60 (1.34) | 3.49 (1.13) | 3.54 (1.22) |
| A5. My instructors have integrated AI generators like ChatGPT into their instruction | 2.19 (1.22) | 2.78 (1.20) | 2.53 (1.24) |
| A6. I plan to use ChatGPT or similar tools for my coursework in the future | 2.66 (1.31) | 4.39 (0.76) | 3.66 (1.34) |
| A7. I have received instructions about how to use ChatGPT or similar tools | 2.30 (1.34) | 3.07 (1.28) | 2.75 (1.36) |
| A8. I would be open to receiving instructions on the use of ChatGPT and/or similar tools | 3.89 (1.11) | 4.36 (0.81) | 4.16 (0.98) |

**Table 8.** Mean and standard deviation of AI perception-related statements for UU, IITD, and overall.

| Statement | UU | IITD | Overall |
|---|---|---|---|
| P1. Students' use of AI text generation tools to complete coursework is prevalent in higher education | 3.63 (1.06) | 4.32 (0.78) | 4.03 (0.97) |
| P2. Students' use of AI text generation tools to complete coursework is inevitable | 3.76 (1.12) | 4.01 (0.98) | 3.90 (1.05) |
| P3. Artificial intelligence has value in education | 3.77 (1.10) | 4.36 (0.82) | 4.11 (0.99) |
| P4. Students should not be restricted from using AI for coursework [a] | 2.89 (1.29) | 3.30 (1.20) | 3.13 (1.25) |
| P5. AI is used in education for good and helpful reasons | 3.54 (1.01) | 3.97 (0.89) | 3.79 (0.97) |
| P6. Instructors are confident in their use of AI in academic settings | 2.69 (0.98) | 3.07 (1.06) | 2.91 (1.04) |
| P7. I would feel confident knowing an instructor was using an AI-created teaching resource | 2.73 (1.18) | 3.27 (1.20) | 3.04 (1.22) |
| P8. I would not need to be informed if my instructor was using AI-created resources on courses [a] | 1.79 (1.01) | 1.89 (0.90) | 1.84 (0.95) |
| P9. I trust AI in marking my assignments and assessments for my modules instead of my instructor | 1.86 (1.05) | 2.55 (1.22) | 2.26 (1.20) |
| P10. Use of AI text generation tools to complete coursework is consistent with academic integrity policies at the university [a] | 2.28 (1.09) | 2.47 (1.03) | 2.39 (1.06) |

[a] Reversed question; for wording as administered, see Table 3.

#### 4.3.1. Exploratory Factor Analysis (EFA)

Exploratory Factor Analysis (EFA) of the 18 items (A1–A8 and P1–P10) was undertaken in MPLus (Muthén & Muthén, 1998). One-, two-, three-, four-, and five-factor solutions were examined. Given the non-independent (clustered in two institutions) nature of the data, the Type = COMPLEX maximum likelihood estimation was used, with a Geomin Oblique rotation.

Loadings for the one-factor solution are presented in Table 9, while there were substantive cross-loadings observed in the multifactor solutions (Appendix B). Therefore, it was decided to base all further analysis on a 15-item, one-factor solution, with items A4, P8, and P10 eliminated (loadings < 0.30). The measure, therefore, is a positive assessment of attitudes towards AI, with a higher score indicative of more positive attitudes.

**Table 9.** Loadings for EFA 1 factor solution. Significant loadings are in bold, with * for $p < 0.05$.

| Item | One |
|---|---|
| A1 | **0.407** * |
| A2 | **0.613** * |
| A3 | **0.783** * |
| A4 | 0.221 * |
| A5 | **0.434** * |
| A6 | **0.827** * |
| A7 | **0.501** * |
| A8 | **0.639** * |
| P1 | **0.594** * |
| P2 | **0.460** * |
| P3 | **0.733** * |
| P4 | **0.480** * |
| P5 | **0.650** * |
| P6 | **0.347** * |
| P7 | **0.515** * |
| P8 | −0.047 |
| P9 | **0.402** * |
| P10 | 0.168 |

Next, we turn to the third research objective to compare the participants' responses across subject areas and institutions.

4.3.2. Interactions Across Subject Areas and Institutions

In order to examine significant between-group differences, we computed a univariate general linear model in SPSS (v.29) with mean AI attitudes score as the dependent variable, and country, discipline, and gender (male or female only[6]) entered as fixed factors. The adjusted $R^2$ for the model was 0.240.

The model revealed statistically significant effects of institution ($F = 83.80$, $p < 0.001$) and gender ($F = 4.88$, $p = 0.027$), but not of subject area ($F = 0.842$, $p = 0.471$). In addition, there was a significant interaction between institution and subject area ($F = 5.471$, $p < 0.001$), suggesting that the main effect of institution may be driven by a particular subject area or areas.

To explore the source of the interaction between institution and subject area, we present participants' attitudes by these two factors in Figure 4 (where error bars represent 95% confidence intervals (CI)). As can be seen, there were no overlapping CIs except in the case of Computer Sciences. The interaction between institution and subject area can therefore be attributed to this contrasting effect for the Computer Sciences subject area: while the main effect of institution is observed for the other subject areas (Biomedical Sciences, Social Sciences, and Technological Sciences), this effect is not observed for Computer Sciences. Rather, Computer Sciences students gave the same AI attitudes ratings across institutions. This contrast is striking, given the specific topic of AI and the specific subject area of Computer Sciences, and shows that the main effect of institution requires a more nuanced interpretation—in particular, one which accounts for subject area.

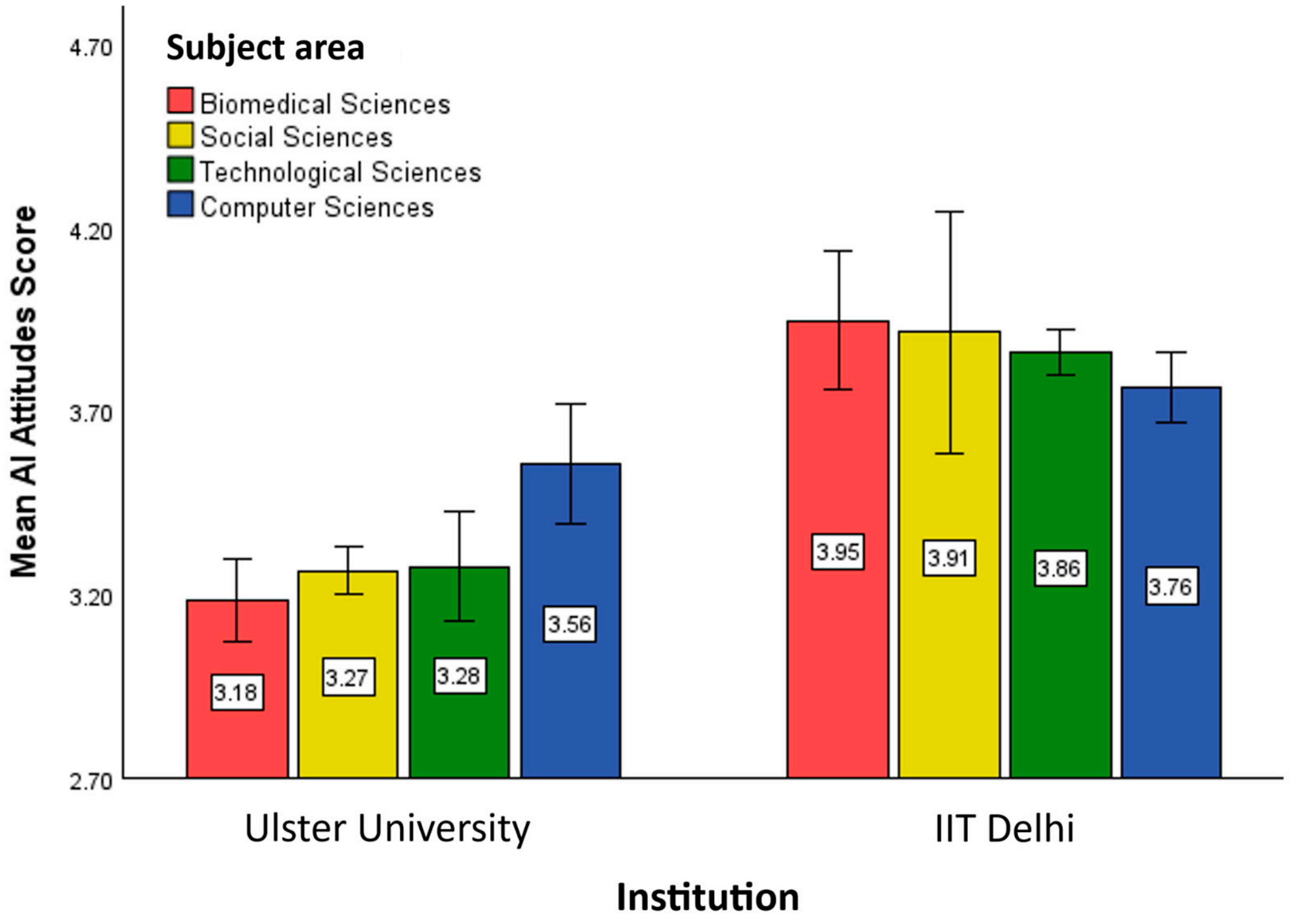

**Figure 4.** Mean AI attitudes scores (institution by subject area).

Meanwhile, there were no other significant two-way interactions: subject area and gender: F = 0.663, $p = 0.575$; institution and gender: F = 0.072, $p = 0.789$; nor was there a significant three-way interaction between institution, subject area and gender (F = 2.415, $p = 0.065$).

This analysis across institutions required a comparison of only those subject areas which were represented at both institutions (cf., Table 4)—that is, participants in the Arts subject area from UU were not included in this analysis (as no Arts students were present in the IITD data), while participants in the Physical Sciences from IITD were not included in this analysis (as no Physical Sciences students were present in the UU data). Therefore, to gain a more complete understanding of the interaction by institution, we focus next on the effects at each institution individually. With this more specific approach, we gain more context across all factors of interest, with the full set of subject areas for each respective institution.

### 4.3.3. Analysis for UU

To investigate the source of the effects of subject area and gender presented in the previous section, the analyses were repeated for the UU sample only, which also included students reporting Arts as their main discipline (Figure 5). The adjusted $R^2$ for the UU model was 0.049. As for the main model with both institutions, we again observed a statistically significant difference in AI attitudes by subject area (F = 5.801, $p < 0.001$). However, the effect of gender was not significant (F = 2.531, $p = 0.112$) in contrast with the main model (which did include a significant effect of gender). We focus first on the effect of subject area, before addressing the contrast in gender effects.

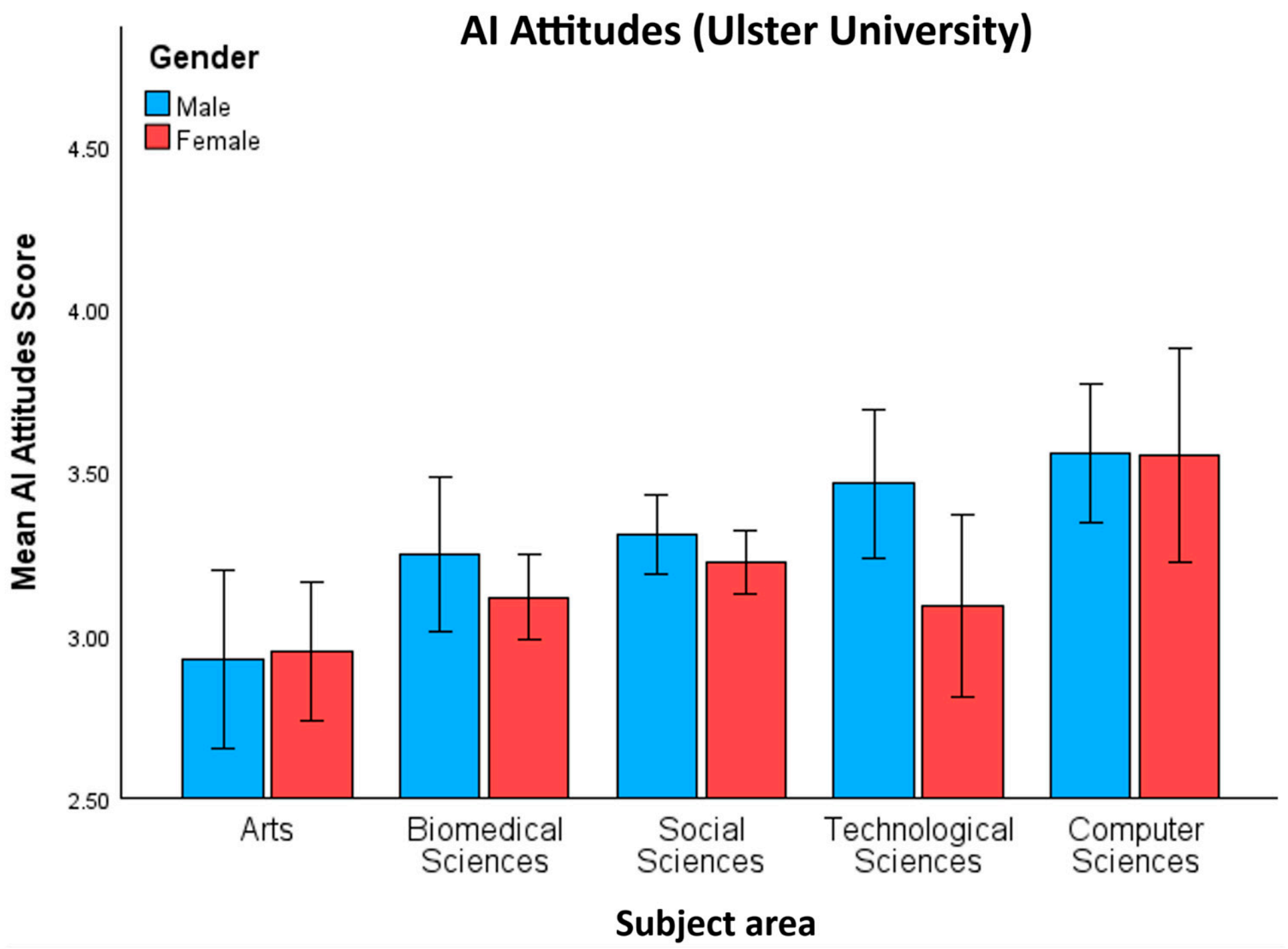

**Figure 5.** AI attitudes for Ulster University, by subject area and gender.

UU: significant effect of subject area

The main effect indicates *that* AI attitudes varied by subject area, but not *how* AI attitudes varied by subject area. This variation is illustrated in Figure 5, by both subject area and gender. As shown in the main analysis, we observe the highest overall attitude scores for participants at UU Computer Sciences. Meanwhile, the scores are numerically lowest for UU participants in Arts subject areas. To identify significant differences between subject areas, we conducted Bonferroni pairwise post hoc tests between each subject area, in Table 10.

**Table 10.** Results of pair-wise Bonferroni post hoc analyses. Significant results are in bold, with * for $p < 0.05$, ** for $p < 0.01$, and *** for $p < 0.001$.

| (A) Science | (B) Science | Mean Diff. (A-B) | Std. Error | Sig. | 95% CI Lower Bound | Upper Bound |
|---|---|---|---|---|---|---|
| Arts | Biomedical | −0.206 | 0.103 | 0.460 | −0.496 | 0.084 |
| | Social | **−0.316 **** | **0.093** | **0.007** | **−0.579** | −0.053 |
| | Technological | **−0.373 *** | **0.124** | **0.027** | **−0.722** | −0.024 |
| | Computer | **−0.616 ***** | **0.124** | **<0.001** | **−0.967** | −0.266 |
| Biomedical | Arts | 0.206 | 0.103 | 0.460 | −0.084 | 0.496 |
| | Social | −0.110 | 0.070 | 1.00 | −0.308 | 0.088 |
| | Technological | −0.167 | 0.107 | 1.00 | −0.470 | 0.136 |
| | Computer | **−0.411 **** | **0.108** | **0.002** | **−0.716** | −0.105 |
| Social | Arts | **0.316 **** | **0.093** | **0.007** | **0.053** | 0.579 |
| | Biomedical | 0.110 | 0.070 | 1.00 | −0.088 | 0.308 |
| | Technological | −0.057 | 0.098 | 1.00 | −0.334 | 0.220 |
| | Computer | **−0.300 *** | **0.099** | **0.026** | **−0.580** | −0.021 |
| Technological | Arts | **0.373 *** | **0.124** | **0.027** | **0.024** | 0.722 |
| | Biomedical | 0.167 | 0.107 | 1.00 | −0.136 | 0.470 |
| | Social | 0.0568 | 0.098 | 1.00 | −0.220 | 0.334 |
| | Computer | −0.244 | 0.128 | 0.581 | −0.605 | 0.118 |
| Computer | Arts | **0.616 ***** | **0.124** | **<0.001** | **0.266** | 0.967 |
| | Biomedical | **0.411 **** | **0.108** | **0.002** | **0.105** | 0.716 |
| | Social | **0.300 *** | **0.099** | **0.026** | **0.021** | 0.580 |
| | Technological | 0.244 | 0.128 | 0.581 | −0.118 | 0.605 |

For the most part, the post hoc tests confirmed the numerical trends in Figure 5: for Arts participants, there were significantly lower attitude scores compared to nearly all other subject areas, aligning with the sentiment analysis presented above. Also as expected, the attitude scores for participants in Computer Sciences were significantly higher than all courses except for Technological Sciences, reflecting the contrast from the original model.

UU: no significant effect of gender

In the main model presented in the previous section, there was a significant effect of gender on AI attitude scores. However, this effect was absent from the model with data from UU only. One consideration for the contrast between the main model and the UU model is the additional Arts subject area in the UU model. However, for the UU model, there was no significant interaction between gender and discipline (F = 0.763, $p = 0.550$). This lack of an interaction strongly suggests that the main model's effect of gender is indeed absent from the UU data, across all subject areas. That is, while no effect of gender was observed for the Arts subject area, this was also the case for the other subject areas (i.e., Biomedical Sciences, Social Sciences, Technological Sciences, and Computer Sciences), which were represented in the main model.

A further consideration for the contrast in the effect of gender would be if the effect in the main model were driven by a gender contrast in the IITD data (discussed further in the following section). However, this would predict a significant interaction in the main model between institution and gender, which was not observed.

The lack of a gender effect in the UU model highlights an important contrast between the mean attitude scores on the one hand, and the distribution of participants across subject areas (cf. Table 4): while the distribution of participants was skewed across the different subject areas with respect to gender, the actual participant ratings did not vary by gender *within* each subject area. Thus, accounting for subject area is a critical step when comparing attitudes across factors which vary by subject area (e.g., gender)[7]. We revisit this contrast in the following section in the context of the IITD data.

#### 4.3.4. Analysis for IITD

Just as for the UU data, we repeated the analysis by subject area and gender for the IITD data, which included ratings from participants in the Physical Sciences, who were not included in the original analysis. The adjusted $R^2$ for this model was 0.007.

Notably—in contrast with both the main model and UU model—the model with IITD data revealed no statistically significant results for subject area (F = 1.637, $p = 0.163$). That is, while ratings in the UU data were modulated by subject area, the ratings from IITD were uniform across subject areas. However, as Table 4 shows, counts for some subject areas were very low, which might prevent any subject-related effects that may exist from appearing to any significant degree. Nevertheless, these ratings were consistently higher across subject areas than those observed in the UU data—with the exception of UU Computer Sciences participants—reflecting the main model's significant effect of institution. These attitude ratings for IITD are illustrated in Figure 6, by subject area and gender.

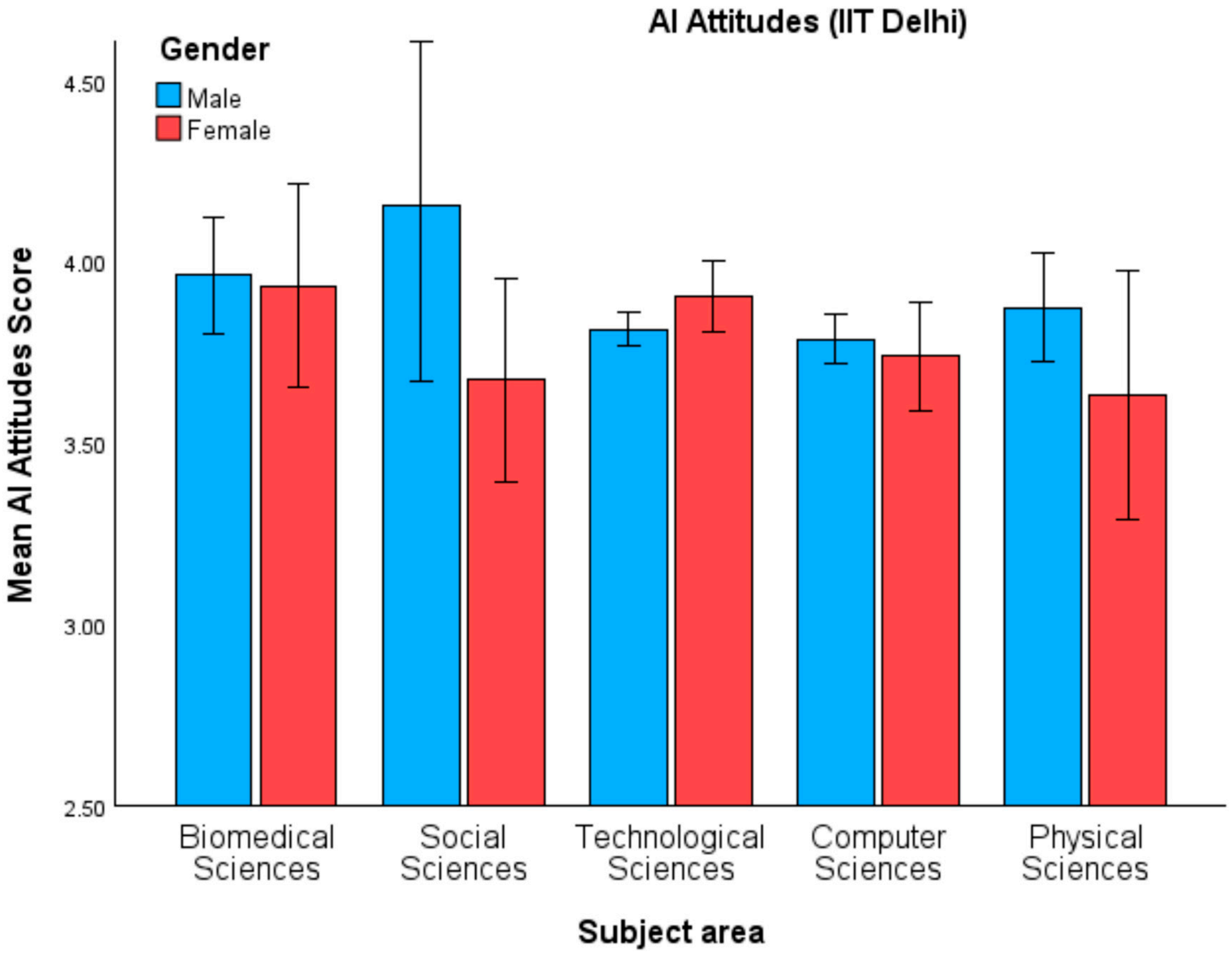

**Figure 6.** AI attitudes for IIT Delhi, by subject area and gender.

## 5. Discussion

In this study, we explored students' perceptions of AI in education contexts and factors that predict variation in perceptions (i.e., differences in perceptions between the different groups). In keeping with previous studies, we observed variation across subject areas, and this variation was reflected both in students' qualitative comments and in responses about their perceptions (Baek et al., 2024; Petricini et al., 2024). We also observed variation across demographic factors in the quantitative responses, particularly participant gender. However, we demonstrated that this effect was an artifact of uneven distributions across subject areas and institutions. Below, we discuss the implications of these results and some considerations for future directions.

### *5.1. Qualitative Data*

For text analysis, we examined both overall and aspect-based sentiment labels. The overall sentiment labels revealed variations in how AI tools are perceived across institutions and academic fields. There was a mix of positive and negative sentiments for UU, while IITD had relatively fewer negative sentiment tags (Table 6). This may reflect that AI tools, like ChatGPT, are viewed as useful for academic writing where English is not a native language—in our case, India. To understand these sentiments in more detail, we explored various aspects and their associated sentiments. With the use of ABSA (see Figures 1 and 2), two major perceptions were revealed: (1) usefulness of AI technology in academic tasks and (2) potential threats with expanding AI technologies.

We have seen that the positive perceptions were associated with aspects such as "questionnaire", "education", "workload", "efficiency", "grammar", etc. On the other side, negative perceptions indicated the issues like "plagiarism", "quality", "performance", "environment", etc. This suggests that students are aware of the broad applications and risks of using AI. Students recognize the usefulness of AI for tasks like ChatGPT improving writing quality; it is also evident as we could see "ChatGPT" in multiple wordclouds in Figures 1 and 2. Also, at the same time, participants are mindful of the risks of AI like academic dishonesty (e.g., "plagiarism"), and environmental costs because of carbon footprints. Other potential inaccuracies can include hallucinations, even though these aspects were not identified from the comments with the ABSA model.

Finally, the correlation heatmap in Figure 3 supports the effectiveness of survey questions to capture general awareness and perceptions; however, the additional optional comments enriched our finding by offering a broader and more variety of perspectives.

### *5.2. Quantitative Data*

To analyze the quantitative ratings on participants' awareness and perceptions, we first compared these responses with the qualitative comments and then conducted an Exploratory Factor Analysis (EFA) to analyze the variation in responses. For the comparison with the qualitative comments, there was some correspondence with the sentiment scores. However, the correlations were stronger between the questions themselves. Similarly, the EFA converged on a 1-factor model, suggesting that participants' responses were largely driven by a single construct. Crucially, this factor was predicted not just by participant demographics, but also by institution and subject area. While previous studies have investigated these factors in isolation, or even in combination as main effects, the current study's analysis included an interaction between these predicting factors.

### *5.3. Predicting Factors*

#### 5.3.1. Interactions

For the single factor identified with the EFA, we explored the effects of institution, subject area, and demographics, focusing on participant gender. Importantly, in addition to the main effects of each variable, we also investigated the interactions between them. These interactions proved crucial for the interpretation of all three variables.

First, we observed a main effect of institution, due to overall higher ratings from students at IITD compared to UU. However, institution also interacted with subject area: this contrast between institutions was not observed for Computer Sciences, for which the same ratings were observed across institutions. This interaction is particularly noteworthy given the central focus on AI, and provides a starting point for further research on variation in students' perceptions (discussed further below).

Second, we observed a main effect of gender, due to overall higher ratings from male students than female students. However, as mentioned above, this effect was an artifact of the distribution of genders across institutions and subject areas: the effect of gender did not persist after inspecting the subject populations at each institution individually, by subject area. This reflects the spurious effects that can emerge with uneven distributions, discussed further in the following section. By focusing on the specific student populations, we control more closely for prior experience with GenAI—a variable which has predicted perceptions in previous studies (e.g., Amoozadeh et al., 2024).

Before turning to the role of experience, we note an important limitation of the analysis by gender, it does not include the full dataset. In particular, the EFA required sufficient responses across each category, and as a result, we could not include non-binary participants or those who did not report their gender in the EFA. This omission is especially important to note, however, given that the ratings from these participants tended to be lower than the ratings that were included. This contrast may reflect bias from various sources in relation to GenAI and must be a priority in future research (Skorodinsky, 2024).

#### 5.3.2. Experience

In previous studies, students' perceptions have been predicted by experience with GenAI, and a similar experience-based variable may explain the results of this study (Faruk et al., 2023; Amoozadeh et al., 2024; Baek et al., 2024). A key consideration, however, is the effects which emerged from the Exploratory Factor Analysis (EFA): ratings differed across subject areas at UU, and overall between UU and IITD. If these effects are driven by different experiences with GenAI, then we would expect these differences to reflect meaningful contrasts across these factors; for example, studying different subject areas at UU would *cause* students to have different experiences with GenAI—with the greatest differences in experience between students in Arts and students in Computer Sciences (see Table 10).

However, the operationalization of "experience" is key. One possibility is that experience differs quantitatively across subject areas at UU, such that students who gave higher ratings have more experience with GenAI. However, a more nuanced interpretation also considers the variation in aspects observed based on the ASBA analysis: comments varied widely across subject areas, therefore indicating differences in qualitative content above and beyond any quantitative contrasts. These differences align with varying influences of GenAI across professional fields, further reflecting that a "one-size-fits-all" approach is insufficient for AI adoption in HE (Li et al., 2024; Baek et al., 2024; Smolansky et al., 2023; see also Kelly et al., 2022). That is, while students desire guidance on best practices for GenAI, this guidance must be tailored to at least some degree (JISC, 2024). For example, this tailoring may be based on prior experience with GenAI use, but also based on the specific topics that this use may be applied to, within different subject areas. While current guidance from HE institutions has tended to target a more general student audience, future guidance may be advised to consider a more tailored approach (e.g., Department for Education, 2023; Ofsted, 2025).The main effect of institution provides further context—both for the variation in students' ratings and for sources of this variation. We observed that ratings were higher for students at IITD than for students at UU across all subject areas, except—crucially—for Computer Sciences, which was matched across the two institutions. This interaction between institution and subject area may shed light on the similarities and differences across institutions. For example, one possible source for the main effect of institution could be if there are specific differences between IITD and UU in the respective subject areas—e.g., specific differences between Biological Sciences at IITD and Biological Sciences at UU, etc. However, this explanation requires an exception for Computer Sciences. Alternatively, the main effect of institution is due to broader differences between UU and IITD—a more plausible possibility given the stronger technology-based focus at IITD.

### *5.4. Limitations and Recommendations*

This study investigated students' perceptions across a range of factors, with a relatively large sample size—1211 students—distributed across the different conditions. Nevertheless, the representativeness of this sample may be improved in future studies, both within and across conditions.

First, while there were relatively large samples across both institutions, these were distributed unevenly across subject areas and demographic variables. Notably, the skew reflected broader variation in how these factors are distributed at both institutions, supporting our sampling methods in a more general sense. However, the variation also reflects biases that may be perpetuated in skewed contexts, due to unequal representation.

Thus, by mirroring a broader population that has unequal distributions, we under-represent the perceptions of participants from minority conditions. This limitation highlights a key consideration when recruiting a "representative" sample: that is, if a sample is representative of the broader population by virtue of matching a relevant distribution, then the sample is less representative of minority sub-groups. Therefore, a "representative" sample may alternatively consist of more even distributions across conditions in order to accurately represent all relevant sub-groups. We recommend this second approach in future studies on perceptions of AI in HE.

## 6. Conclusions

This study assessed the awareness and perceptions of AI in higher education across borders, and explored these measures through the lenses of two major categories—namely, institutions and subject area. The study was based on a large-scale survey with over 1K participants from UU and IITD, representing the global western vs. eastern world, developed vs. developing, and native vs. non-native English contexts, respectively. Importantly, we observed variation in perceptions across both categories, and this variation was observed in both the qualitative comments and quantitative responses. Crucially, we found that the variation was further modulated by interactions between institution and subject area, which accounted for apparent variation by demographic factors. These results constitute a key contribution to the literature as they highlight important considerations for interpreting variation in the context of multiple factors, and in future studies we will further explore these interacting factors and their implications for GenAI in higher education.

In addition, the analysis indicated that the variation across the institutions is primarily due to variation in major focus area, as well as the technical nature of the IITD in comparison to UU. We also observed the usefulness of optional text comments in better assessing the perception, for example, by enlisting the use cases as well as potential risks to better map their awareness and perception around GenAI. Finally, we found that the responses were correlated for many of the questions, which was further clarified by showing that the participants' responses were largely driven by a single construct/factor.

**Author Contributions:** Conceptualization, J.G., M.S., M.M. (Morgan Macleod), A.R.; methodology, J.G., M.M. (Morgan Macleod), A.R.; software, M.M. (Morgan Macleod); validation, J.G., M.S., M.M. (Morgan Macleod); formal analysis, S.S., M.M. (Morgan Macleod), M.M. (Michael McKay); investigation, J.G., M.M. (Morgan Macleod); data curation, M.M. (Morgan Macleod), S.S.; writing—original draft, J.G., S.S., M.M. (Morgan Macleod), M.M. (Michael McKay), M.S.; writing—review and editing, M.S., T.C.; visualization, S.S., M.M. (Michael McKay); supervision, J.G., M.S., T.C.; project administration, J.G.; funding acquisition, J.G., M.S., T.C. All authors have read and agreed to the published version of the manuscript.

**Funding:** This work has received funding from the International Science Partnerships Fund (ISPF) managed by the Northern Ireland Department for Science, Innovation and Technology.

**Institutional Review Board Statement:** The study was conducted in accordance with the Declaration of Helsinki, and approved by the Ethics Committee of the Ulster University School of Communication and Media, CMFC-24-004 and February 2024.

**Informed Consent Statement:** Informed consent was obtained from all subjects involved in the study.

**Data Availability Statement:** Dataset available on request from the authors.

**Acknowledgments:** We are thankful to the ISPF project team for support in developing the questionnaire, including Kelly Norwood (Ulster University), Muhammad Usman Hadi (Ulster University), Aisling Reid (Queen's University Belfast), Zhiwei Lin (Queen's University Belfast), Jocelyn Dautel (Queen's University Belfast), Adina Camelia Bleotu (University of Bucharest) and Caitlin Meyer (University of Amsterdam). We would also like to thank Sue Attewell (JISC) and Nicola Bartholomew (Ulster University), as well as the attendees of the workshop on Adaptive Education: Harnessing AI for Academic Progress for helpful discussion. Last but not least, we greatly appreciate the time spent by the survey participants from IIT Delhi and Ulster University.

**Conflicts of Interest:** The authors declare no conflicts of interest.

## Abbreviations

The following abbreviations are used in this manuscript:

| | |
|---|---|
| UU | Ulster University |
| IITD | Indian Institute of Technology Delhi |
| ASBA | aspect-based sentiment analysis |
| EFA | Exploratory Factor Analysis |

## Appendix A

**Table A1.** Demographics questions.

| Question | Choices |
|---|---|
| What is your course? | Free text |
| What school (department) are you in? | Select from list of schools |
| Are you an undergraduate or postgraduate student? | Undergraduate<br>Postgraduate |
| What is your current year group? | First year<br>Second year<br>Placement year<br>Final year<br>Other |
| Which of the following best describes your gender identity? | Male<br>Female<br>Non-binary<br>Other |
| Which of the following best describes your age range? | 18–24<br>25–39<br>40–59<br>60+ |
| What is your ethnic group? (Northern Ireland questionnaire) (select multiple if mixed) | White<br>Irish Traveller<br>Indian<br>Chinese<br>Roma<br>Filipino<br>Black African<br>Black Other<br>Other |
| What is your religion? (India questionnaire) (select multiple if mixed) | Hindu<br>Islamic<br>Sikh<br>Christian<br>Other Religion |

# Appendix B

**Table A2.** Two-factor solution. Significant loadings are in bold, with * for $p < 0.05$.

| Item | Factor 1 | Factor 2 |
|---|---|---|
| A1 | **0.847** * | 0.749 * |
| A2 | **0.923** * | 0.751 * |
| A3 | **0.691** * | **0.355** * |
| A4 | 0.217 * | **0.302** * |
| A5 | −0.007 | **0.555** * |
| A6 | 0.355 | **0.653** * |
| A7 | 0.073 * | **0.553** * |
| A8 | 0.352 | 0.424 |
| P1 | 0.435 | **0.305** * |
| P2 | 0.344 | 0.219 |
| P3 | 0.369 | **0.519** |
| P4 | 0.209 | 0.356 |
| P5 | 0.180 | **0.593** * |
| P6 | −0.114 | **0.537** * |
| P7 | 0.051 | **0.664** * |
| P8 | 0.262 * | 0.188 * |
| P9 | 0.186 | **0.669** * |
| P10 | 0.009 | 0.204 |

**Table A3.** Three-factor solution. Significant loadings are in bold, with * for $p < 0.05$.

| Item | Factor 1 | Factor 2 | Factor 3 |
|---|---|---|---|
| A1 | **0.847** * | 0.749 * | 0.530 * |
| A2 | **0.923** * | 0.751 * | 0.019 |
| A3 | **0.691** * | **0.355** * | **0.400** * |
| A4 | 0.217 * | **0.302** * | −0.018 |
| A5 | −0.007 | **0.555** * | −0.017 |
| A6 | 0.355 | **0.653** * | −0.008 |
| A7 | 0.073 * | **0.553** * | 0.011 |
| A8 | 0.352 | 0.424 | 0.052 |
| P1 | 0.435 | **0.305** * | 0.172 * |
| P2 | 0.344 | 0.219 | 0.123 * |
| P3 | 0.369 | **0.519** | −0.007 |
| P4 | 0.209 | 0.356 | −0.116 * |
| P5 | 0.180 | **0.593** * | −0.148 * |
| P6 | −0.114 | **0.537** * | −0.165 |
| P7 | 0.051 | **0.664** * | −0.282 |
| P8 | 0.262 * | 0.188 * | −0.246 * |
| P9 | 0.186 | **0.669** * | −0.328 * |
| P10 | 0.009 | 0.204 | −0.138 |

**Table A4.** Four-factor solution. Significant loadings are in bold, with * for $p < 0.05$.

| Item | Factor 1 | Factor 2 | Factor 3 | Factor 4 |
|---|---|---|---|---|
| A1 | **0.847** * | 0.749 * | **0.530** * | −0.045 |
| A2 | **0.923** * | 0.751 * | 0.019 | 0.859 * |
| A3 | **0.691** * | **0.355** * | **0.400** * | 0.803 * |
| A4 | 0.217 * | **0.302** * | −0.018 | 0.497 * |
| A5 | −0.007 | **0.555** * | −0.017 | 0.013 |
| A6 | 0.355 | **0.653** * | −0.008 | 0.789 * |
| A7 | 0.073 * | **0.553** * | 0.011 | 0.184 * |
| A8 | 0.352 | 0.424 | 0.052 | **0.678** * |

**Table A4.** *Cont.*

| Item | Factor 1 | Factor 2 | Factor 3 | Factor 4 |
|---|---|---|---|---|
| P1 | 0.435 | **0.305** * | 0.172 * | 0.638 * |
| P2 | 0.344 | 0.219 | 0.123 * | **0.534** * |
| P3 | 0.369 | **0.519** | −0.007 | **0.845** * |
| P4 | 0.209 | 0.356 | −0.116 * | 0.638 * |
| P5 | 0.180 | **0.593** * | −0.148 * | **0.653** * |
| P6 | −0.114 | **0.537** * | −0.165 | 0.051 |
| P7 | 0.051 | **0.664** * | −0.282 | 0.424 * |
| P8 | 0.262 * | 0.188 * | −0.246 * | 0.127 * |
| P9 | 0.186 | **0.669** * | −0.328 * | 0.222 * |
| P10 | 0.009 | 0.204 | −0.138 | 0.222 * |

**Table A5.** Five-factor solution. Significant loadings are in bold, with * for $p < 0.05$.

| Item | Factor 1 | Factor 2 | Factor 3 | Factor 4 | Factor 5 |
|---|---|---|---|---|---|
| A1 | **0.847** * | −0.180 * | **0.749** * | **0.530** * | −0.045 |
| A2 | **0.923** * | 0.000 | **0.751** * | 0.701 | 0.019 |
| A3 | **0.691** * | **0.355** * | **0.400** * | **0.803** * | 0.048 |
| A4 | 0.217 * | 0.097 | **0.302** * | −0.018 | **0.497** * |
| A5 | −0.007 | **0.555** * | −0.017 | 0.013 | **0.838** * |
| A6 | 0.355 | **0.653** * | −0.008 | **0.789** * | 0.076 |
| A7 | 0.073 * | **0.553** * | 0.011 | 0.184 * | **0.619** * |
| A8 | 0.352 | 0.424 | 0.052 | **0.678** * | −0.062 |
| P1 | 0.435 | **0.305** * | 0.172 * | **0.638** * | −0.058 |
| P2 | 0.344 | 0.219 | 0.123 * | **0.534** * | −0.125 * |
| P3 | 0.369 | **0.519** | −0.007 | **0.845** * | −0.169 * |
| P4 | 0.209 | 0.356 | −0.116 * | **0.638** * | −0.268 |
| P5 | 0.180 | **0.593** * | −0.148 * | **0.653** * | 0.005 |
| P6 | −0.114 | **0.537** * | −0.165 | 0.051 | **0.546** * |
| P7 | 0.051 | **0.664** * | −0.282 | **0.424** * | 0.177 |
| P8 | **0.262** * | 0.188 * | −0.246 * | 0.127 * | 0.140 * |
| P9 | 0.186 | **0.669** * | −0.328 * | 0.222 * | 0.338 * |
| P10 | 0.009 | 0.204 | −0.138 | 0.222 * | −0.081 |

## Notes

1 We note that the survey questions used in this study refer to "AI" rather than "GenAI" specifically; however, students' comments often conflated these terms, primarily in using "AI" to mean "GenAI", and this practice can also be observed in previous studies on perceptions. Thus, while we review previous studies on GenAI perceptions, our methods maintain the original study wording of "AI".

2 The postgraduate category included students in degree programs after an undergraduate degree, generally Level 7–8; the category did not include doctoral students.

3 https://huggingface.co/cardiffnlp/twitter-roberta-base-sentiment-latest

4 https://huggingface.co/tomaarsen/setfit-absa-paraphrase-mpnet-base-v2-restaurants-aspect

5 https://github.com/amueller/word_cloud

6 Due to small numbers reported for the non-binary and other categories.

7 This consideration is distinct from the more fundamental question of why such factors tend to vary by subject area, which is beyond the scope of this study.